\documentclass[epj,final]{svjour}
\newcommand{\Tr}{{\rm Tr}}

\usepackage{amsmath,amssymb}
\usepackage{graphics}
\begin{document}
\title{
Reflection symmetry in mean--field replica--symmetric spin glasses}
\author{N.V.Gribova\inst{1}, V.N.Ryzhov\inst{1},
T.I.Schelkacheva\inst{1},   \and
E.E.Tareyeva\inst{1}
\thanks{\emph{Present address:} Institute for High Pressure Physics,
Russian Academy of Sciences, Troitsk 142092, Moscow region,
Russia}%
}                     
\offprints{}          
\institute{Institute for High Pressure Physics, Russian Academy of
Sciences, Troitsk 142092, Moscow region, Russia}
\date{Received: date / Revised version: date}
%
\abstract{ The role of reflection symmetry breaking for the
character of the appearance of replica symmetric spin glass state
is investigated. We establish the following symmetry rule for
classical systems with one order parameter in the replica
symmetric mean field approximation. If in the pure system the
transition to the ordered phase is of the second order, then in
the corresponding random system the glass regime appears as a
result of a phase transition; if the transition in the pure system
is of the first order, then glass and ordered
regimes grow continuously in the random system on cooling.
\PACS{
      {64.70.-p, 64.70.Kb}   
      {}{}
     } 
} 
\titlerunning{Reflection symmetry in
mean--field replica--symmetric ...
}

\maketitle

The crucial role of the reflection symmetry for the character of
phase transition in nonrandom mean--field (MF) models is well
known (see, e.g., the textbook ~\cite{LLSt}). Generally speaking
the presence of the terms without reflection symmetry causes the
first order phase transition, while in the absence of such terms
the transition is of the second order. Usually this result is
obtained in the frame of the phenomenological approach based on
the Ginsburg--Landau (GL) effective Hamiltonian that can be easily
obtained for any Hamiltonian through the Hubbard--Stratanovich
identity for the partition function.  On a slightly "more
microscopic level" the statement about the order of phase
transition can be demonstrated considering the MF one--site
equations as is done below (for continuous case see the
Eqs.(11-13) of ~\cite{TMFY}).

Let us consider a system of particles on lattice sites $i, j$ with
Hamiltonian
\begin{gather}
H=-\frac{1}{2}\sum_{i\neq j}J_{ij}
\hat{U_i}\hat{U_j}, \label{one}
\end{gather}
where $\hat{U}$ is a diagonal operator with   $\Tr\hat{ U}=0$,
the interactions $J_{ij}$ being such that the MF approximation
gives exact solution. For example, $J_{ij}$ may be random exchange
interactions with Gaussian probability distribution
 \begin{equation} P(J_{ij})=\frac{1}{\sqrt{2\pi
J}}\exp\left[-\frac{(J_{ij}-J_0)^{2}}{
2J^{2}}\right] \label{two}
\end{equation}
with $ J=\tilde{J}/\sqrt{N}$ , $J_{0}=\tilde{J_0}/N$. In nonrandom
pure case $J=0$. Let us construct GL Hamiltonian as an integral on
fluctuations $\varphi$. Using the Hubbard-Stratanovich identity
the partition sum can be written in the following form:
\begin{multline}
Z=\Tr\exp\left(\frac{1}{2}\beta\sum_{i
j}^N{\hat{U}}_{i}J_{ij}\hat{U}_j\right)=
\\
\frac{{\left[ \det{(\beta
J)}^{-1}\right]}^{1/2}}{{\left[2\pi\right]}^{N/2}}
\int_{-\infty}^{\infty}...\int_{-\infty}^{\infty}
\prod_{i=1}^{N}d
{\varphi}_{i}\times
\\
\exp\left\{-\frac{1}{2}\sum_{ij}{\varphi}_i{(\beta
J)}^{-1}_{ij}{\varphi}_{j}+
\sum_{i}\ln\Tr\exp\left[{\varphi}_i{\hat{U}}_i\right]\right\}=
\\
C\int_{-\infty}^{\infty}...\int_{-\infty}^{\infty}\prod_{i=1}^{N}d
{\varphi}_{i}\exp\left\{-\beta H_{eff}\right\} \label{part}
\end{multline}
The effective Hamiltonian contains the powers $n$ of the
fluctuations $\varphi$ for which $\Tr( \hat{U}^n)$ is nonzero.

When dealing with random MF systems the role of cubic term is
discussed usually in connection with the replica symmetry breaking
(RSB) solution of the equations for glass order parameters.  The
absence of reflection symmetry causing cubic terms in GL
Hamiltonian for regular nonrandom system results in a special form
of RSB free energy functional for random case, too. This form
gives the discontinuity for the order parameter, the stability of
the first stage RSB solution and other specific features (see, for
example, Refs.~\cite{goeld,kirkp,rev3}). However, in this paper we
accent our attention on the replica symmetric (RS) approach. We
investigate the role of the reflection symmetry for the behavior
of RS solution for spin--glass--like MF systems with one regular
order parameter in the Hamiltonian and formulate a kind of
symmetry rule for the type of the growing of glass regime: if in
the nonrandom (pure) system the transition to the ordered phase is
of the second order, then in the corresponding random system the
glass regime appears as a result of a phase transition; if the
transition in the pure system is of the first order, then in the
random system the glass regime grows continuously on cooling. In
fact, one can imagine that both the first order phase transition
in nonrandom systems as well as the continuous growing of the
glass regime in random systems are caused by some kind of internal
fields appearing due to the algebra of operators $\hat U$. Some
previous results can be found in ~\cite{lrst}. It is well known
that a number of magnetic compounds with phase transition to
spin--glass state behaves qualitatively as the
Sherrington--Kirkpatrick (SK) model ~\cite{sk} (without $\varphi
^3$ term). We would like to bring the readers attention to the
fact that the case with the continuous growing of the glass regime
(with $\varphi ^3$ term) also is a reality and exists in some
mixed crystals ~\cite{sul2,lrt}.

Now let us consider the system~(\ref{one})-~(\ref{two}). Using
replica approach (see, e.g. ~\cite{sk}) we can write the
 free energy in the form
\begin{multline}
\langle F\rangle_J/NkT=-\lim_{n \rightarrow
0}\frac{1}{n}\max\biggl \{-\sum_{\alpha}
 \frac{(x^{\alpha})^{2}}{2}-
 \\
\sum_{\alpha}
 \frac{(w^{\alpha})^{2}}{2} - \sum_{\alpha>\beta}
 \frac{(y^{\alpha\beta})^{2}}{2}+
\\
\ln\Tr_{\{U^{\alpha}\}}\exp\biggl[\sum_{\alpha}
 x^{\alpha}\sqrt{\frac{\tilde{J_0}}{kT}}U^\alpha+
 \sum_{\alpha}w^{\alpha}\frac{1}{\sqrt 2}t(U^{\alpha})^2+
 \\
\sum_{\alpha>\beta}y^{\alpha\beta}tU^{\alpha}U^{\beta}\biggr]\biggr\}
 \label{three}
\end{multline}
where
\begin{gather}
 m^{\alpha}=(x^{\alpha})^{\rm extr}/\sqrt{\frac{\tilde{J_0}}{kT}}={\langle U^{\alpha}\rangle}_{\rm
eff};\notag
\\
 q^{\alpha\beta}=(y^{\alpha\beta})^{\rm extr}/t  = {\langle
 U^{\alpha}U^{\beta}\rangle}_{\rm eff};\label{four}
 \\
p^{\alpha}=(w^{\alpha})^{\rm
extr}\sqrt{2}/t={\langle(U^{\alpha})^{2}\rangle}_{\rm eff},\notag
\end{gather}
as follows from the saddle point equations. Here $t=\tilde{J}/kT$
and averaging is performed with the effective Hamiltonian ${\cal
H}_{eff}$:
$$-{\cal H}_{eff}
=\sum_{\alpha}\frac{\tilde{J_0}}{kT}m^{\alpha}U^{\alpha}+
\sum_{\alpha}\frac{t^{2}}{2}p^{\alpha}(U^{\alpha})^{2}+
\sum_{\alpha
>\beta}t^{2}q^{\alpha,\beta}U^{\alpha}U^{\beta}$$

      In the RS approximation the free energy has the form:
\begin{multline}
 F=-NkT\left\{-\left(\frac{\tilde{J_0}}{kT}\right)\frac{m^2}{2}+
 t^2\frac{q^2}{4}-t^2\frac{p^2}{4}+\right.
\\
\left.\int_{-\infty}^{\infty}\frac{dz}{\sqrt{2\pi}}\exp\left(-\frac{z^2}{2}\right)\ln
\Tr\left[exp\left(\hat{\theta}\right)\right]\right\} \label{frs}
\end{multline}
 Here
$$\hat{\theta}=\left[zt\sqrt{q}+m\left(\frac{\tilde{J_0}}{kT}\right
)\right]\hat{U}+t^2\frac{p-q}{2}\hat{U}^2.$$ The order parameters
are: $ m $ is the regular order parameter (an analog of magnetic
moment in spin glasses), $ q$ is the glass order parameter and $p$
is an auxiliary order parameter. The extremum conditions for the
free energy ~(\ref{frs}) give the following equations for these
order parameters:
\begin{gather}
m=\int_{-\infty}^{\infty}\frac{dz}{\sqrt{2\pi}}\exp\left(-\frac{z^2}{2}\right)
\frac{\Tr\left[\hat{U} \exp\left(\hat{\theta}\right)\right]}
{\Tr\left[\exp\left(\hat{\theta}\right)\right]} \label{mrs}
\\
q=\int_{-\infty}^{\infty}\frac{dz}{\sqrt{2\pi}}
\exp\left(-\frac{z^2}{2}\right)
\left\{\frac{\Tr\left[\hat{U}
\exp\left(\hat{\theta}\right)\right]}
{\Tr\left[\exp\left(\hat{\theta}\right)\right]}\right\}^{2}
\label{qrs}
\\
p=\int_{-\infty}^{\infty}\frac{dz}{\sqrt{2\pi}}
\exp\left(-\frac{z^2}{2}\right)
\frac{\Tr\left[\hat{U}^2 \exp\left(\hat{\theta}\right)\right]}
{\Tr\left[\exp\left(\hat{\theta}\right)\right]} \label{prs}
\end{gather}

In general case the order parameter $p$ is not independent from
the others. For Ising spin glass \cite{sk} $\hat{U}=\hat{S}$,
$(\hat{U})^2=(\hat{S})^2=1=p$ and Eq.~(\ref{prs}) reduces to an
identity. For the quadrupolar glass with $J=1$ \cite{lrt} we have
$\hat{U}=\hat{Q}=3{\hat{J_z}}^2-2$,  so that $p=2-m$. However, for
the quadrupolar glass with $J=2$ ~\cite{qg2} $p$ is independent
order parameter: the subalgebra contains the operators $\hat
{Q}$,
$(\hat{Q})^2$ and the unit matrix $\hat{E}$, so that only
$(\hat{Q})^3$ is not independent and can be represented as a
linear combination of $\hat{Q}$, $(\hat{Q})^2$ and $\hat{E}$. In
the case of $S=1$ spin glass \cite{spin1} we have $S_z=0,\pm1$,
$p$ is independent parameter and has the physical meaning defined
by the equality ${\hat{S}}^2=1/3(2+\hat{Q})$.  Now
$p=(1/3)(2+m_2)$, where $m_2$ is the quadrupolar regular order
parameter.

In the random case $J_0=0$ the high temperature expansion of the
equations (\ref{mrs})  --  (\ref{prs}) has the form:
\begin{gather}
m = \frac {t^2}{2}A_3 p + \frac{t^4}{8}p^2(A_5 - 2 A_2 A_3) -
\frac{t^4}{2} A_2 A_3 q^2 -\notag
\\
-\frac{t^4}{2}A_2 A_3 p q + ... \label{mht}
\\
q = t^2 A_2^2 q + \frac{t^4}{2} A_3^2 p^2 + \frac{t^4}{2} q^2
(A_3^2 - 4 A_2^3) + t^4 A_2 A_4 qp + ... \label{qht}
\\
p = A_2 + \frac{t^2}{2}p (A_4 - A_2^2) + \frac{t^4}{8}p^2 (A_6 -
3A_2A_4 + 2 A_2^3)-\notag
\\
-\frac{t^4}{2}q^2A_2(A-4 - A_2^2) - \frac{t^4}{2}pqA_3^2 + ...
\label{pht}
\end{gather}
If $m=0,q=0$ and $p\neq0$ then
\begin{gather}
\Tr[\hat{U}\exp(\hat{\theta})]=\Tr[\hat{U}\exp(t^2p{\hat{U}}^2/2)]
\end{gather}
is a sum of $ A_n=\Tr({\hat{U}}^n)$ with odd $n$ only. So if
$A_{(2k+1)}=0$ then Eqs.(\ref{mrs})  -- (\ref{prs}) have the
solution $m=0,q=0$. If $A_{(2k+1)}\neq0$ Eqs.(\ref{mrs})  --
(\ref{prs}) have no trivial solution $m=0$, $q=0$ at any
temperature.  The high temperature expansion gives explicitly
$m\sim{t}^2A_3, q\sim{t}^4 A_3^2$. The glass order parameter grows
continuously on cooling.

In the case $A_{(2k+1)}=0$ one can obtain from (\ref{qrs}) the
bifurcation point $T_c^{b}$ where the nontrivial solution  for the
glass order parameter $q$ appears under the condition $m~=~0$. The
equations for $T_c^{b}$ have the form:
$$ p(t_c) = 1/t_c;~~~~t_c\equiv \tilde{J}/kT_c^b $$
$$ p(t_c) =
\Tr[\hat{U^2}\exp(t^2p(t_c){\hat{U}}^2/2)].$$

The replica symmetric solution is stable unless the replicon
mode energy $\lambda$ is nonzero. For our model we have:
\begin{multline}
\lambda_{repl} = 2 - 2 t^2
\int_{-\infty}^{\infty}\frac{dz}{\sqrt{2\pi}}\exp\left(-\frac{z^2}{2}\right)
\left\{\frac{\Tr\left[\hat{U}^2
\exp\left(\hat{\theta}\right)\right]}
{\Tr\left[\exp\left(\hat{\theta}\right)\right]}-\right.
\\
\left.\left[\frac{\Tr\left[\hat{U}
\exp\left(\hat{\theta}\right)\right]}
{\Tr\left[\exp\left(\hat{\theta}\right)\right]}\right]^2\right\}^2.
\label{lambda}
\end{multline}
The character of the replica symmetry breaking is defined by the
term $\alpha_3 (\delta q_{\alpha\beta})^3$ in the RSB free
energy (\ref{three}) expansion  near RS free energy (\ref{frs}).
We have
$$\alpha_3= - t^6 \int_{-\infty}^{\infty}\frac{dz}{\sqrt{2\pi}}\exp\left(-\frac{z^2}{2}\right)
\left[\frac{\Tr\left[\hat{U}^3
\exp\left(\hat{\theta}\right)\right]}
{\Tr\left[\exp\left(\hat{\theta}\right)\right]}\right]^2$$
$$-3 t^6 q
\int_{-\infty}^{\infty}\frac{dz}{\sqrt{2\pi}}\exp\left(-\frac{z^2}{2}\right)
\left[\frac{\Tr\left[\hat{U}^2
\exp\left(\hat{\theta}\right)\right]}
{\Tr\left[\exp\left(\hat{\theta}\right)\right]}\right]^2+2t^6
q^3 .$$

If all $A_{2n+1}=0$ then
$\Tr\left[\hat{U}^3 \exp\left(\hat{\theta}\right)\right]=0$ and
$q=0$ at $T=T_c$ so that $\alpha_3=0$ and at $T_c$ the full
replica symmetry breaking according to the Parisi scheme takes
place. If $A_{2n+1}\neq0$ then $\alpha_3\neq0$ and the first
stage RSB solution is stable.

In the regular case $(\tilde{J}=0,\tilde{J_0}\neq0)$ we have from
Eqs.(\ref{mrs}) -- (\ref{prs}) $q\equiv{m}^2$ and
\begin{equation}
m=\frac{
\Tr\left[\hat{U}
\exp\left(m\left(\frac{\tilde{J_0}}{kT}\right)\hat{U}\right)\right]}
{\Tr\left[\exp\left(m\left(\frac{\tilde{J_0}}{kT}\right)\hat{U}\right)\right]}
\label{mreg}
\end{equation}
\begin{equation}
p=\frac{\Tr\left[{\hat{U}}^2
\exp\left(m\left(\frac{\tilde{J_0}}{kT}\right)\hat{U}\right)\right]}
{\Tr\left[\exp\left(m\left(\frac{\tilde{J_0}}{kT}\right)\hat{U}\right)\right]}
\label{preg}
\end{equation}
These mean field equations have the trivial solution $m_0~=~0$ and
$p_0=\Tr({\hat{U}}^2)/\Tr(\hat{E})$ for all temperatures. A
non-trivial solution for $m$  appears at the bifurcation point
$k{T_c}^b/\tilde{J_0}=p_0$.  Near ${T_c}^b$ the expansion of
(\ref{mreg}) -- (\ref{preg}) gives the bifurcation equation
\begin{multline}
\frac{m^2}{2} A_2 {{\lambda}_0}^2-\tau A_2- \frac{m}{2} A_3
{{\lambda}_0}^2-\frac{m^2}{6} A_4 {{\lambda}_0}^3 -
\\
- m\tau{\lambda}_0 A_3 +...=0 \label{bifreg}
\end{multline}
where $\frac{\tilde{J_0}}{kT}=\lambda,
\frac{\tilde{J_0}}{k{T_c}^b}=\lambda_0, \lambda=\lambda_0+\tau$.

If $A_3=0$, the solution of (\ref{bifreg}) has the form
\begin{equation}
m=\sqrt{\tau \frac{A_2}{ \frac{1}{2} A_2
{{\lambda}_0}^2-\frac{1}{6} A_4 {{\lambda}_0}^3 }}=\sqrt{
\frac{6\tau A_2^4}{3A_2^2-A_4}} \label{mtau}
\end{equation}
and we have the second order phase transition. If $A_3\neq0$ the
solution of Eq.(\ref{mreg}) in the vicinity of $T_c^b$ is
\begin{equation}
 m=-\frac{2\tau A_2^3}{A_3}
 \label{mtau3}
 \end{equation}
In this case the phase transition is of the first order.The
transition temperature can be obtained from the comparison of free
energies of ordered and disordered phases. (see, e.g. Ref.\cite
{TMFY,c60}).

So one can establish the following symmetry rule for classical
systems with one (pure) order parameter in the replica-symmetric
MFA. If in the pure system the transition to the ordered phase is
of the second order, then in the corresponding random system the
glass regime appears as a result of a phase transition; if the
transition in the pure system is of the first order, then in the
random system glass and ordered regimes grow smoothly on cooling.

As an example let us now consider  the Hamiltonian (\ref{one})
with $\hat{U}=\hat{Q}+\eta\hat{V}$, $\hat{Q}=3\hat{J_z}^2-2$,
$\hat{V}=\sqrt{3}({\hat{J_x}}^2-{\hat{J_y}}^2)$, $J = 1$, $J_z =
1, 0, -1$. This model describes random quadrupole interaction. A
molecular quadrupolar moment is the second-rank tensorial operator
with five independent components. In the principal axes frame only
two of them remain: $\hat{Q}$ and $\hat{V}$. It is easy to show
that ${\hat{Q}}^2=2-\hat{Q}$, ${\hat{V}}^2=2+\hat{Q}$,
$\hat{Q}\hat{V}=\hat{V}\hat{Q}=\hat{V}$. Here $\eta>0$ is a tuning
parameter.
\begin{equation}
(\hat{Q}+\eta\hat{V})^2=2(1+\eta^2)+2\eta\hat{V}+(1-\eta^2)\hat{Q}.
\label{alg}
\end{equation}

The corresponding Ginzburg-Landay Hamiltonian written in terms of
fluctuation fields $\varphi_i$ has the form:
\begin{gather}
\beta H_{eff}=\frac{1}{2}\sum_{ij}{\varphi}_i{(\beta
J)}^{-1}_{ij}{\varphi}_{j}-N\ln3-\label{GL}
\\
N\sum_{i} \left[{\varphi}_i^2(1+{\eta}^2)+
{\varphi}_i^3({\eta}^2-1/3)
+{\varphi}_i^4{(1+{\eta}^2)}^2/4+...\right]\notag
\end{gather}

The RS free energy has the form
\begin{multline}
F=-NkT\left\{-\left(\frac{\tilde{J_0}}{kT}\right)\frac{m_1^2}{2}+
t^2\frac{q^2}{4}-\right.
\\
\left.-t^2\frac{m_2^2}{4}+ t^2{(1+\eta^2)}^2-t^2(1+\eta^2)q+
+\right.
\\
\left.+\int_{-\infty}^{\infty}\frac{dz}{\sqrt{2\pi}}\exp\left(-\frac{z^2}{2}\right)
\ln\psi\right\} \label{fqv}
\end{multline}
Here
$$\psi=\exp(-2\vartheta_1)+\exp(\vartheta_1)\left[\exp(\vartheta_2)+
\exp(-\vartheta_2)\right]$$
$${\vartheta}_1=zt\sqrt{q}+m_1\left(\frac{\tilde{J_0}}{kT}\right)
+t^2({\eta}^2-1)\left[({\eta}^2+1)-\frac{m_2}{2}-\frac{q}{2}\right]$$
$${\vartheta}_2=\eta\sqrt{3}t^2[2(1+{\eta}^2)-(m_2+q)]+
\eta\sqrt{3}\left[m_1\left(\frac{\tilde{J_0}}{kT}\right)+
zt\sqrt{q}\right].$$ The order parameters are
$m_1\sim\ll\hat{Q}+\eta\hat{V}\gg$, $q$ is the corresponding glass
order parameter, the order parameter
$m_2\sim\ll(1-\eta^2)\hat{Q}-2\eta\hat{V}\gg$ appears due to the
operators algebra (see Eq.~(\ref{alg})) in close analogy with
 the
appearing of quadrupole order parameter in the case of spin glass
with $S=1$. We use $m_2$ instead of $p$: $$p=2({\eta}^2+1)-m_2.$$

The equations for the order parameters have the form:
\begin{equation}
m_1=\int_{-\infty}^{\infty}\frac{dz}{\sqrt{2\pi}}\exp\left(-\frac{z^2}{2}\right)
\left[\frac{\frac{\partial\psi}{\partial{\vartheta}_1}+
\eta\sqrt{3}\frac{\partial\psi}{\partial{\vartheta}_2}}{\psi}\right],
\label{m1qv}
\end{equation}
\begin{equation}
m_2=-\int_{-\infty}^{\infty}\frac{dz}{\sqrt{2\pi}}\exp\left(-\frac{z^2}{2}\right)
\left[\frac{({\eta}^2-1)\frac{\partial\psi}{\partial{\vartheta}_1}+
2\eta\sqrt{3}\frac{\partial\psi}{\partial{\vartheta}_2}}{\psi}\right],
\label{m2qv}
\end{equation}
\begin{equation}
q=\int_{-\infty}^{\infty}\frac{dz}{\sqrt{2\pi}}\exp\left(-\frac{z^2}{2}\right)
\left[\frac{\frac{\partial\psi}{\partial{\vartheta}_1}+
\eta\sqrt{3}\frac{\partial\psi}{\partial{\vartheta}_2}}{\psi}\right]^2,
\label{qqv}
\end{equation}

It is easy to see that
$$(\hat{Q}+\eta\hat{V})^3=2(3\eta^2-1)+3(1+\eta^2)(\hat{Q}+\eta\hat{V})];$$
$$\Tr(\hat{Q}+\eta\hat{V})^3 = 6(3\eta^2-1)$$
Using these equalities and Eqs. ~(\ref{GL}) one can show that the
behavior of the system under consideration is quite different in
the cases $\eta = 1/\sqrt{3}$ and $\eta\neq 1/\sqrt{3}$.

Let us consider first the regular case ($\tilde{J}=0$,
$\tilde{J_0}\neq0$). Now we have from
Eqs.~(\ref{mreg})--~(\ref{mtau3}):
$$m_1 = \frac{16}{3}\sqrt{\tau} , 3\eta^2=1;$$
$$m_1 = - \frac{8 \tau (1+\eta^2)^3}{3\eta^2-1},
3\eta^2-1\neq0.$$ The case $\eta = 1/\sqrt{3}$ has transparent
physical meaning for $\tilde{J_0}>0$, second order transition for
spin one ferromagnet being an example. In the case
$3\eta^2-1\neq0$ we obtain the first order phase transition for
$\eta>1/\sqrt{3}$ and $\tilde{J_0}>0$ or for $\eta<1/\sqrt{3}$ and
$\tilde{J_0}<0$. It should be noted that an example of such a
behavior is presented by the case $\eta=0$ describing the
orientational quadrupolar ordering in $o-H_2$ or $p-D_2$ (see
~\cite{TMFY,hydrogen}).  We can  re
write equation (\ref{mreg}) as
\begin{equation}
\Phi(y,\lambda)=(y-4\lambda)\exp(3y/2)+2y+4\lambda=0 \label{fi}
\end{equation}
where $y=-2m\left(\frac{\tilde{J_0}}{kT}\right)$,
$\lambda=\left(\frac{\tilde{J_0}}{kT}\right)$.

The bifurcation points $\lambda_i$ for $\Phi(y,\lambda)$
\cite{hydrogen} are obtained from the condition
$\left(\frac{\partial\Phi}{\partial y}\right)=0$. We have
$$ \lambda_1=1/2;y_1=0;m_(1)=0;$$
$$\lambda_2=1/2.18;y_2=0.7;m_(2)=-0.8.$$

The form of the solutions can be obtained from the expansion of
(\ref{fi}) in powers of $\tau=\lambda-{\lambda}_\alpha$ and
$\xi=y-y_\alpha$. Because $\Tr{\hat{Q}}^3\neq0$ in the vicinity of
${T_1}^b=\tilde{J_0}/k\lambda_1$ we have $\xi\sim\tau$. However in
the vicinity of ${T_2}^b=1.09{T_1}^b$ we have
$\xi\sim\sqrt{\tau}$. The physical solution near ${T_2}^b$ takes
place only for $\mu>0 , T<{T_2}^b$. It is a turning point. The
sharp phase transition with the jump of $m$ is between ${T_1}^b$
and ${T_2}^b$.  The transition point is calculated from the
comparison of energies. Other cases with $\eta\neq1/\sqrt{3}$ can
be described in a similar way.
\begin{figure}[htb]
\resizebox{0.5\textwidth}{!}{%
  \includegraphics{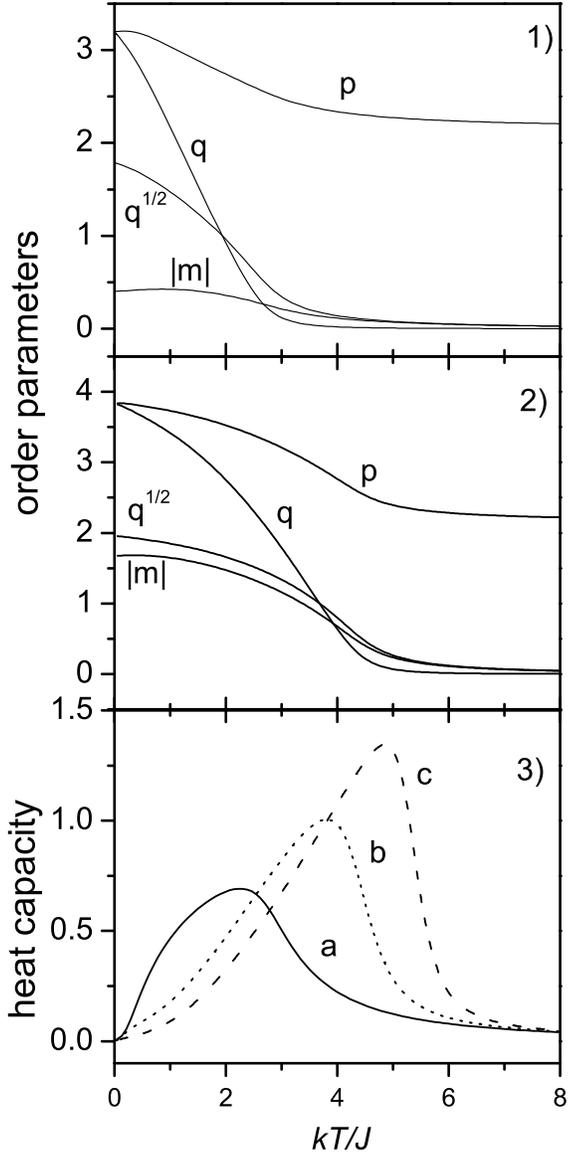}}
\caption{Order parameters for $\eta=0.5/\sqrt{3}$ 1) $\tilde{J}_0
= 0$; 2) $\tilde{J}_0/\tilde{J} = 1.5$; and 3)  specific heat for
a)$\tilde{J}_0/\tilde{J} = 0$; b)$\tilde{J}_0/\tilde{J} = 1.5$;
c)$\tilde{J}_0/\tilde{J} = 2$.} \label{fig:1}
\end{figure}

\begin{figure}[htb]
\resizebox{0.5\textwidth}{!}{\includegraphics{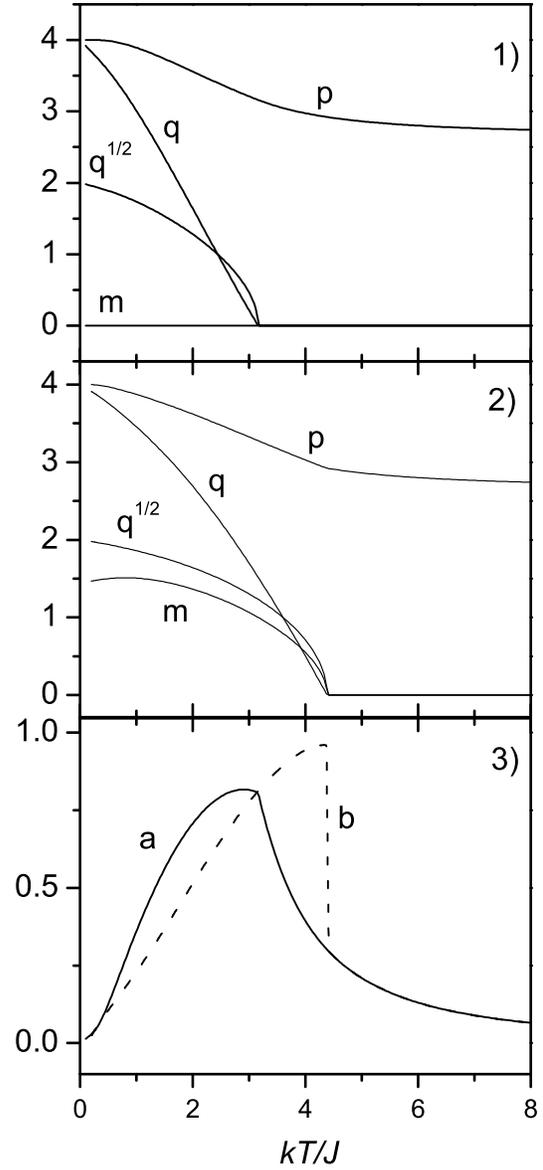}}
\caption{Order parameters for $\eta=1/\sqrt{3}$ 1) $\tilde{J}_0 =
0$; 2) $\tilde{J}_0/\tilde{J} = 1.5$; and 3)  specific heat for
a)$\tilde{J}_0/\tilde{J} = 0$ ; b)$\tilde{J}_0/\tilde{J} = 1.5.$}
\label{fig:2}
\end{figure}

Let us consider now the random case
($\tilde{J}\neq0$,$\tilde{J_0}=0$). The first terms of the high
temperature expansion of Eqs.~(\ref{mrs}) -- ~(\ref{prs}) in the
case $\hat{U} =\hat{Q}+\eta \hat{V}$ can be easily obtained from
~(\ref{mht}) -- ~(\ref{pht}) and have the form:
\begin{gather}
m_1 = 2t^2ab + t^4ab^3 - t^2am_2 - t^4m_2ab^2 + \frac{t^4}{4}m_2ab
- \notag\\
-4t^4ab^2q - 2t^4abq^2 + 2t^4abm_2q, \label{mrsht}
\\
q = 4t^2 b^2 q + 4t^4 a^2 b^2 - 4t^4 ba^2m_2\notag
\\
+ t^4 a^2 m_2^2 + 8 t^4b^4q - 4t^4 b^3 qm_2 + 2t^4 q^2 (a^2 -
8b^2), \label{qrsht}
\\
m_2 = -2b^3t^2 + t^2b^2m_2 - t^4b^2(2a^2 - b^2) + t^4bm_2(2a^2 -
b^2)\notag
\\
- \frac{t^4}{4}(2a^2 - b^2)m_2^2 + 2t^4b^3q^2 + 4t^4a^2bq -
2t^4a^2m_2q. \label{prsht}
\end{gather}
Here
$$a = 3\eta^2 - 1,~~~~~~~b = 1 + \eta^2.$$

If $\eta=1/\sqrt{3}$ the Hamiltonian  becomes analogous to the
$S=1$ spin glass Hamiltonian. There is
the trivial solution $q=0$
when $T>{T_c}^b$ because $A_3=0$. The solution $q\neq0$ appears at
the point ${kT_c}^b/\tilde{J}\approx3.2$ (for $m=0$) \cite{spin1}.
The specific heat function loses smoothness at this point (see
fig.1). In the case $a\neq0$ there is no trivial solution of
Eq.~(\ref{qrsht}) and $q$, $m_2$ and $m_1$ grows continuously on
cooling. This case is analogous to quadrupolar glass with two
internal fields. The absence of the zero solution at high
temperatures for glass order parameter is unambiguously connected
with the absence of reflection symmetry.

In the figures we present the behavior of the order parameters for
different cases as well as the specific heat
\begin{multline}
\frac{C_v}{kN}=\frac{d}{d(kT/\tilde{J})}
\left\{\left(\frac{\tilde{J}}{kT}\right)\frac{(q^2-p^2}{2}\right\}
-
\\
-\left(\frac{\tilde{J_0}}{\tilde{J}}\right)m_1
\frac{dm_1}{d(kT/\tilde{J})}
   \label{heat}
\end{multline}

Let us note that the symmetry rule for the type of the growing of
glass regime and phase transition in pure systems have been
obtained in MFA. It is interesting to notice that in the Bethe
approximation (and in some other cluster approximations) the
internal fields appear in the case of nonzero cubic terms and
cause a smooth changing of the order parameter without phase
transition in nonrandom system if in MFA the phase transition is
of the first order (see, e.g., ~\cite{etters}). In the systems
with reflection symmetry and the MFA second order phase transition
the rather sharp phase transition from nonzero order parameter
phase remains in the Bethe approximation. In some way the replica
approach to random systems effects analogously to Bethe
approximation.

The simpliest way to demonstrate this fact is to consider cluster
approximation of ~\cite{clust}. The main equation is the
selfconsistency condition given in terms of effective mean field
$\psi$ produced by the neighbors and acting on the particle on
each site of the cluster.

For the Hamiltonian ~(\ref{one}) in nonrandom case with nearest
neighbor interaction we obtain the following equations ~\cite{cl,clust}:
\begin{multline}
<\hat{U}>=\frac{\Tr\hat{U_1}\exp\left[-\beta
\hat{H_1}(\psi)\right]}{\Tr\exp\left[-\beta
\hat{H_1}(\psi)\right]}=
\\
\frac{\Tr\hat{U_1}\exp\left[-\beta
\hat{H_s}(\psi)\right]}{\Tr\exp\left[-\beta
\hat{H_s}(\psi)\right]} \label{clust}
\end{multline}
where $s$ is the number of particles in the cluster. For the
simplest case $s=2$ we have where
$H_1=-\hat{U_1}\psi;H_2=-J\hat{U_1}\hat{U_2}
-\psi\left(1-1/\gamma\right)\left(\hat{U_1}+\hat{U_2}\right)$;
$\hat{U_i}$ is the operator $\hat{U}$ on $i$-th site; $\gamma$ is
the number of nearest neighbors. The effective mean field  $\psi$
can be obtained from ~(\ref{clust}). The high temperature
expansion for ~(\ref{clust}) is
\begin{multline}
\label{cl1}\psi \beta J \Tr\left({\hat{U}}^2\right)+{\psi}^2
\frac{{\beta}^2J^2}{2}\Tr\left({\hat{U}}^3\right)=\psi \alpha
\beta\Tr\left({\hat{U}}^2\right)+
\\
\frac{{\beta}^2J^2}{2}\Tr\left({\hat{U}}^2\right)
\Tr\left({\hat{U}}^3\right)+
{\psi}^2\frac{{\alpha}^2{\beta}^2}{2}\Tr\left({\hat{U}}^3\right)+
\\
+\psi {\beta}^2\alpha{\left[\Tr\left({\hat{U}}^2\right)\right]}^2,
\end{multline}
where $\alpha=1-1/\gamma$.

It is easy to see from ~(\ref{cl1}) that there is no solution
$\psi~=~0$ for the case $\Tr\left({\hat{U}}^3\right)\neq0$.  So
there is no phase transition and the order parameter grows
smoothly on cooling. When $\Tr\left({\hat{U}}^3\right)=0$ a
bifurcation point can exist for Eq.~(\ref{cl1}) and in the pure
system we obtain a phase transition.

In conclusion we formulate a kind of symmetry rule for the way of
the appearance of glass regime: if in the nonrandom (pure) system
the transition to the ordered phase is of the second order, then
in the corresponding random system the glass regime appears as a
result of a phase transition; if the transition in the pure system
is of the first order, then in the random system the glass regime
grows continuously on cooling \cite{sul2}.

This work was supported in part by the Russian Foundation for
Basic Research (Grants 02-02-16621 (N.V.G. and E.E.T.) and
02-02-16622 (V.N.R. and T.I.S.)).
%
%

\end{document}